\documentclass[namedreferences]{SolarPhysics}
\usepackage[optionalrh]{spr-sola-addons} 
\usepackage{graphicx}        
\usepackage{color}           
\usepackage{url}             
\usepackage{lscape}
\usepackage{longtable}
\usepackage{multirow}

\usepackage{amsmath}
\usepackage{float}

\newcommand{\etal}{{\it et al.}}

\newcommand{\aap}{    {\it Astron. Astrophys.}}
\newcommand{\aaps}{   {\it Astron. Astrophys. Suppl.}}

\newcommand{\apj}{    {\it Astrophys. J.}}

\newcommand{\cjaa}{   {\it Chin. J. Astron. Astrophys.}}

\newcommand{\jgr}{    {\it J. Geophys. Res.}}

\newcommand{\solphys}{{\it Solar Phys.}}

\newcommand{\ssr}{    {\it Space Sci. Rev.}}
\newcommand{\phil}{    {\it Phil. Trans. R. Soc. London}}
\newcommand{\jcp}{    {\it J. Comput. Phys.}}

\begin{document}

\begin{article}

\begin{opening}

\title{Comparison of Nonlinear Force-free Field and
Potential Field in the Quiet Sun\\ {\it Solar Physics}}

\author{S. Liu$^{1}$\sep
       H.Q. Zhang$^{1}$\sep
        J.T. Su$^{1}$
       }
\runningauthor{S. Liu et al.} \runningtitle{Comparison of Nonlinear
Force-free Field and Potential Field in the Quiet Sun}

   \institute{Key Laboratory of Solar Activity,
   National Astronomical Observatory, Chinese Academy of Sciences,
        Beijing, China
                     email: \url{lius@nao.cas.cn}\\
             }

\begin{abstract}
In this paper, a potential field extrapolation and three nonlinear
force-free (NLFF) field extrapolations (optimization, direct
boundary integral (DBIE) and approximate vertical integration (AVI)
methods) are used to study the spatial configuration of magnetic
field in the quiet Sun. It is found that the strength differences
between the three NLFF and potential fields exist in the low layers.
However, they tend to disappear as the height increases, which are of
the order of 0.1 G when the height exceeds $\sim$ 2000 km above the
photosphere. The absolute azimuth difference between one NLFF field
and the potential field is as follows: for the optimization field,
it decreases evidently as the height increases; for the DBIE field,
it almost keeps constant and shows no significant change as the
height increases; for the AVI field, it increases slowly as the
height increases. The analysis shows that the reconstructed NLFF
fields deviate significantly from the potential field in the quiet
Sun.
\end{abstract}
\end{opening}
\section{Introduction}
     \label{S-Introduction}
Although a large amount of studies focused on the magnetic field of
solar active regions, it is very important to know the spatial
configuration of magnetic field in the quiet Sun since most part of
solar surface are covered by the quiet regions even in the solar
maximum years. Due to the restrictions of observational technique,
the accurate information about how the magnetic field transport in
the solar chromosphere and corona is still not clear. Theoretically,
\citeauthor{gab76} (\citeyear{gab76}) predicted that the magnetic
field expands from the photosphere into the chromosphere and corona
to form a canopy-like structure. At present, it is still difficult
to check \citeauthor{gab76}'s suggestion from the observations. In
spite of that, the relations between the photospheric magnetic field
and the chromospheric and coronal features may provide some
important but limited information about the expansion of magnetic
field.

Some authors studied the relations between the photospheric and
chromospheric magnetic field ($i.g.,$ \citeauthor {zhang96},
\citeyear{zhang96}; \citeauthor{alme97}, \citeyear{alme97};
\citeauthor{demou97}, \citeyear{demou97}; \citeauthor{cupe97},
\citeyear{cupe97}; \citeauthor{zhang99}, \citeyear {zhang99};
\citeauthor{zhangh00}, \citeyear{zhangh00}; \citeauthor{zhang00},
\citeyear{zhang00} and \citeauthor{harvey06a},
\citeyear{harvey06a}). For example, \citeauthor{zhang96}
(\citeyear{zhang96}) argued that the magnetic field of the active
regions extends up into the chromosphere in the fibril forming from
the photosphere. \citeauthor{zhangh00} (\citeyear{zhangh00}) found
there are similarities between the chromospheric and photospheric
magnetograms in the quiet Sun. Some other authors studied the
relations between the magnetic fields in the photosphere and
transition regions/corona ($i.g.,$ \citeauthor{giova80}
\citeyear{giova80}; \citeauthor{dow86} \citeyear{dow86};
\citeauthor{sch03} \citeyear{sch03}; \citeauthor{harvey06b}
\citeyear{harvey06b}; \citeauthor{philip08} \citeyear{philip08} and
\citeauthor{dim09} \citeyear{dim09}). For example,
\citeauthor{sch03} (\citeyear{sch03}) argued that relatively strong
internetwork field lines close back within several thousand
kilometers. \citeauthor{dim09} (\citeyear{dim09}) found that there
are no direction correlations between the fractal dimensions of the
2D photospheric patterns and their 3D counterparts in the corona at
the nonlinear force-free limit, but there are significant
correlations between the fractal dimensions of the photospheric and
coronal structures for the potential and linear force-free (LFF)
extrapolation.

The magnetic field extrapolation with force-free assumption
(\citeauthor{aly89}, \citeyear{aly89}) is an alternative method to
study the configuration of the solar magnetic field based on the
observations of photospheric magnetic field. The merit is that it
can give the adequate spatial information of the solar magnetic
field. Generally, most of the extrapolated fields have been used to
describe the topology of magnetic field in the active regions
($e.g.,$ \citeauthor{reg02}, \citeyear{reg02}; \citeauthor{reg04},
\citeyear{reg04}; \citeauthor{wie06},\citeyear{wie06};
\citeauthor{song06}, \citeyear{song06}; \citeauthor{son07},
\citeyear{son07}; \citeauthor{he08a}, \citeyear{he08a};
\citeauthor{sch08}, \citeyear{sch08}; \citeauthor{jing08},
\citeyear{jing08} and \citeauthor{der09}, \citeyear{der09}).
However, as measurements of the vector magnetic fields of active
regions are more reliable than those of the quiet Sun ($e.g.,$
better signal-to-noise ratio in strong field areas). Moreover,
force-free assumption is even more widely violated in the quiet Sun
than in the active regions, therefore so far very few studies
applied extrapolation techniques to study the configuration of
magnetic field in the quiet Sun. \citeauthor{regnier08}
(\citeyear{regnier08}) used the magnetic field extrapolation to
derive the null points density in the quiet Sun. \citeauthor{tu05}
(\citeyear{tu05}) used the magnetic field extrapolation to study the
origin of solar wind and give the 3D structure of magnetic field in
a quiet Sun. They found that there is no canopy shape as originally
suggested, but the cross section of magnetic flux increases almost
linearly with height. Their study is consistent with the results by
\citeauthor{zhang00} (\citeyear{zhang00}). However, the method used
in above studies is the potential field approximation or LFF field
extrapolation. Since the magnetic field in the quiet Sun is
non-potential (\citeauthor{wood99}, \citeyear{wood99};
\citeauthor{zhao09}, \citeyear{zhao09}), use of the NLFF field
extrapolation is more reasonable than that of the linear force-free
extrapolation. Therefore, the difference between the NLFF field and
the potential field in the quiet Sun is an important subject to
study.

At present, the potential and LFF field extrapolations reached a
mature development. In this case, only the line-of-sight (LOS)
component of the magnetic field is taken as the boundary condition
($e.g.,$ \citeauthor{chiu77}, \citeyear{chiu77};
\citeauthor{seehafer78}, \citeyear{seehafer78}; \citeauthor{alis81},
\citeyear{alis81}; \citeauthor{gary89}, \citeyear{gary89}). For the
NLFF field extrapolation, which, on the contrary, requires the
knowledge of the vector field at the photosphere, recently several
models and methods have been proposed ($e.g.,$, \citeauthor{wu90},
\citeyear{wu90}; \citeauthor{ama97}, \citeyear{ama97};
\citeauthor{sak81}, \citeyear{sak81}; \citeauthor{cho81},
\citeyear{cho81}; \citeauthor{yan00}, \citeyear{yan00};
\citeauthor{whe00}, \citeyear{whe00}; \citeauthor{thomas04},
\citeyear{thomas04}; \citeauthor{song06}, \citeyear{song06};
\citeauthor{he08}, \citeyear{he08}). Most of these methods can give
reliable results that satisfy force-free assumption ($e.g.,$
\citeauthor{sch06}, \citeyear{sch06}; \citeauthor{ama06},
\citeyear{ama06}; \citeauthor{song06}, \citeyear{song06};
\citeauthor{valo07}, \citeyear{valo07}). In applications to the
measured magnetograms from the active regions, the reliability of
the extrapolated fields is often assessed by checking the
morphological consistence between the NLFF model field lines and the
observed features such as coronal loops observed in EUV and X-ray
images ($e.g.$, \citeauthor{reg04}, \citeyear{reg04};
\citeauthor{wie06}, \citeyear{wie06} and \citeauthor{reg07},
\citeyear{reg07}). However, at present there is no such work that
uses the NLFF field extrapolation to study the configuration of
magnetic field in the quiet Sun, mainly because the vector
magnetograms in the quiet Sun before the {\it Hinode} were not very
suitable for the application of NLFF extrapolation. Fortunately,
Spectro-Polarimeter (SP) of the Solar Optical Telescope (SOT) on
board {\it Hinode} (\citeauthor{kosugi07}, \citeyear{kosugi07};
\citeauthor{tsuneta08}, \citeyear{tsuneta08};
\citeauthor{ichimoto08}, \citeyear{ichimoto08}) has been used to
measure the vector magnetic field in the quiet Sun with very high
spatial resolution and adequate sensitivity for the first time. This
gives us a chance to extrapolate the NLFF magnetic field in the
quiet Sun.

Organization of this paper is as follows: firstly, the description
of employed data and the extrapolation methods will be introduced in
Section~\ref{S-Data}, secondly, the comparisons between the NLFF
fields and the potential field are shown in Section~\ref{S-Result},
at last, the discussions and conclusions will be given in
section~\ref{S-Discussion}.

\section{Data processing and extrapolation methods}
    \label{S-Data}
\subsection{Data processing}
A quiet region observed by {\it Hinode} SOT/SP on April 16, 2007
from 00:23 UT to 01:48 UT is used in this work. The SP obtains line
profiles of two magnetically sensitive Fe lines at 630.15 and 630.25
nm and nearby continuum. Spectra are exposed and read out
continuously 16 times per rotation of the polarization modulator,
and the raw spectra are added and subtracted onboard in real time to
demodulate, generating Stokes $IQUV$ spectral images. The parameters
relevant to the vector magnetic field, which are derived from the
inversion of the full Stokes profiles based on the assumption of the
Milne-Eddington (ME) atmospheric model, are the total field strength
B, the inclination angle $\gamma$ with respect to the (LOS)
direction, the azimuth angle $\phi$ and the filling factor $f$.
Following \citeauthor{lit99} (\citeyear{lit99}) and
\citeauthor{zhao09} (\citeyear{zhao09}), the longitudinal component
of the spatially resolved vector field is obtained with the
expression $fBcos(\gamma)$, and the transverse component
$\sqrt{f}Bsin(\gamma)$. We use the acute angle method
(\citeauthor{wang94}, \citeyear{wang94}; \citeauthor{wang97},
\citeyear{wang97}; \citeauthor{wang01}, \citeyear{wang01} and
\citeauthor{mat06}, \citeyear{mat06}) to resolve $180^{\circ}$
ambiguity, in which the observed field is compared to the
extrapolated potential field at the photosphere. The orientation of
the observed transverse component is chosen by requiring that
$-90^{\circ}$ $\leq$ $\bigtriangleup \theta$ $\leq$ $90^{\circ}$,
where $\bigtriangleup \theta$ = $\theta_{o}$-$\theta_{e}$ is the
angle between the observed and extrapolated transverse components.

Figure 1A shows the LOS magnetogram of this quiet region observed by
{\it Hinode} SOT/SP. Figure 1B is the vector magnetogram that is
employed as the extrapolation boundary in this work, and it
corresponds to the region that is highlighted by a white square in
Figure 1A. The size of the employed vector magnetogram is 100
$\times$ 100 pixels with a resolution of 0.148$^{''}$ in x-direction
and 0.159$^{''}$ in y-direction. It is located near the solar disk
center (-5.2$^{''}$ and 7.3$^{''}$ in X-Y direction of the
heliographic coordinates). In our work the altitude of extrapolated
fields is limited in 50 pixels and its resolution is 0.148$^{''}$ in
z-direction.

\subsection{Four extrapolation methods}
The force-free assumption requires the magnetic field to satisfy
the following equations,
\begin{equation}
\label{free1}
\nabla\times \textbf{B} = \alpha(\textbf{r}) \textbf{B},
\end{equation}
\begin{equation}
\label{free2}
\nabla\cdot \textbf{B} = 0,
\end{equation}
They imply that there is no Lorentz force and $\alpha$ is constant
along magnetic field lines. If $\alpha$ = 0, the equations represent
a potential field (a current-free field). If $\alpha$ = constant,
they describe a current-carrying LFF field; if $\alpha$ = $f(\textbf{r})$
they describe a general NLFF field.

In this paper, the LFF extrapolation method
(\citeauthor{seehafer78}, \citeyear{seehafer78}) is used to
calculate the potential field by choosing $\alpha$ = 0. This method
gives the components of the magnetic field in terms of a Fourier
series. The photospheric magnetogram that covers a region with a
length of $L_{x}$ in x-direction and a length of $L_{y}$ in
y-direction is artificially extended to a rectangular region
covering -$L_{x}$ to $L_{x}$ and -$L_{y}$ to $L_{y}$ by taking an
antisymmetric mirror image of the original magnetogram. For example,
$B_{z}(-x,y)$ = $-B_{z}(x,y)$ and $B_{z}(x,-y)$ = $-B_{z}(x,y)$. The
expression for the magnetic field is given by

\begin{flalign}
\begin{split}
 B_{x} = \sum_{m,n=1}^{\infty}\dfrac{C_{mn}}{\lambda_{mn}}{\rm exp}(-r_{mn}z)
[\alpha\dfrac{n\pi}{L_{y}}\sin(\dfrac{m\pi x}{L_{x}})\cos(\dfrac{n\pi y}{L_{y}})\\
- r_{mn}\dfrac{m\pi}{L_{x}}\cos(\dfrac{m\pi x}{L_{x}})\sin(\dfrac{n\pi y}{L_{y}})],
  \hspace{18ex}
\end{split}
\end{flalign}
\begin{flalign}
\begin{split}
 B_{y} = - \sum_{m,n=1}^{\infty}\dfrac{C_{mn}}{\lambda_{mn}}{\rm exp}(-r_{mn}z)
[\alpha\dfrac{m\pi}{L_{x}}\cos(\dfrac{m\pi x}{L_{x}})\sin(\dfrac{n\pi y}{L_{y}}) \\
+ r_{mn}\dfrac{n\pi}{L_{y}}\sin(\dfrac{m\pi x}{L_{x}})\cos(\dfrac{n\pi y}{L_{y}})],
  \hspace{19ex}
\end{split}
\end{flalign}
\begin{flalign}
\begin{split}
B_{z} = \sum_{m,n=1}^{\infty} C_{mn}{\rm exp}(-r_{mn}z)
\sin(\dfrac{m\pi x}{L_{x}})\sin(\dfrac{n\pi y}{L_{y}}),
\end{split}
\end{flalign}
with $\lambda_{mn}$ = $\pi^{2}(m^{2}/L^{2}_{x} + n^{2}/L^{2}_{y})$
and $r_{mn}$ = $\sqrt{\lambda_{mn} - \alpha^{2}}$. The coefficients
$C_{mn}$ are obtained by taking the FFT of $B_{z}$ at z = 0.

The optimization method presented by \citeauthor{whe00}
(\citeyear{whe00}) and developed by \citeauthor{thomas04}
(\citeyear{thomas04}) consists in minimizing a joint measure for the
normalized Lorentz force and the divergence of the field, given by the
function,
\begin{equation}
\label{opit} L = \int_{V}\omega(x,y,z)[B^{-2}|(\nabla \times
\textbf{B}\times \textbf{B}) |^{2}+|\nabla\cdot
\textbf{B}|^{2}]d^{3}x,
\end{equation}
where $\omega(x,y,z)$ is a weighting function. It is evident that
(for $w > 0$) the force-free equations are fulfilled when $L$ is
equal to zero. This method involves minimizing $L$ by optimizing the
solution function $\textbf{B}(x, t)$ through states that are
increasingly force- and divergence-free, where $t$ is an artificial
time-like parameter. The relevant theories and algorithms can be
found in the papers of \citeauthor{whe00} (\citeyear{whe00}) and
\citeauthor{thomas04} (\citeyear{thomas04}).

The direct boundary integral equation (DBIE) method
(\citeauthor{yan06}, \citeyear{yan06}; \citeauthor{he08},
\citeyear{he08}) is developed from the BIE method proposed by
\citeauthor{yan00} (\citeyear{yan00}), which uses the Green function
to extrapolate the magnetic field. In this method, an optimized
parameter $\lambda$, defined in the papers of \citeauthor{yan06}
(\citeyear{yan06}), must be found through iteration. The integral
\begin{equation}
\label{dbie}
\textbf{B}(x_{i}, y_{i}, z_{i}) =
\int_\Gamma\dfrac{z_{i}[\lambda r \sin(\lambda r)+\cos(\lambda r)]
\textbf{B}_{0}(x, y, 0)} {2\pi[(x-x_{i})^2+(y-y_{i})^2+z_{i}^2]^{3/2}},
\end{equation}
is used to calculate the magnetic field, where $r$ =
$[(x-x_{i})^2+(y-y_{i})^2+z_{i}^2]^{1/2}$ and $\textbf{B}_{0}$ is
the magnetic field of photospheric surface. The detailed theories
can be found in the papers of \citeauthor{he08} (\citeyear{he08})
and \citeauthor{yan00} (\citeyear{yan00}).

The approximate vertical integration (AVI) method
(\citeauthor{song06}, \citeyear{song06}) was improved from the
direct integration proposed by \citeauthor{wu90} (\citeyear{wu90}).
In this method, the magnetic field is given by the following
formula,
\begin{equation}
\label{q-avi1}
\textbf{B}_{x} = \xi_{1}(x,y,z)F_{1}(x,y,z),
\end{equation}
\begin{equation}
\textbf{B}_{y} = \xi_{2}(x,y,z)F_{2}(x,y,z),
\end{equation}
\begin{equation}
\label{q-avi2}
\textbf{B}_{z} = \xi_{3}(x,y,z)F_{3}(x,y,z),
\end{equation}
assuming the second-order continuous partial
derivatives in a certain height range, 0$ <z<$H (H is the calculated
height from the photospheric surface). In Equations
(\ref{q-avi1})-(\ref{q-avi2}), $\xi_{1}, \xi_{2}$ and $\xi_{3}$
mainly depend on $z$ and slowly vary with $x$ and $y$, while $F_{1},
F_{2}$ and $F_{3}$ mainly depend on $x$ and $y$ and weakly vary with
$z$, which are mathematical
representation of the similarity solutions. After constructing the
magnetic field, the following integration equations,

\begin{equation}
\label{a}
 \dfrac{\partial B_{x}}{\partial z} =
 \dfrac{\partial B_{z}}{\partial x} + \alpha B_{y},
\end{equation}
\begin{equation}
 \dfrac{\partial B_{y}}{\partial z} =
 \dfrac{\partial B_{z}}{\partial y} - \alpha B_{x},
\end{equation}
\begin{equation}
 \dfrac{\partial B_{z}}{\partial z} =
 -\dfrac{\partial B_{x}}{\partial x} -
 \dfrac{\partial B_{y}}{\partial y},
\end{equation}
\begin{equation}
\label{b}
 \alpha B_{z} = \dfrac{\partial B_{y}}{\partial x}
  - \dfrac{\partial B_{x}}{\partial y},
\end{equation}
are used to carry out the extrapolation. The detailed descriptions
are described in the paper of \citeauthor{song06}
(\citeyear{song06}).

\section{Results}
\label{S-Result} Conventionally, the quiet Sun is considered far
from force-free, thus the force-free extent of the selected area on
the photosphere should be investigated. Three parameters
$F_{x}/F_{p}$, $F_{y}/F_{p}$ and $F_{z}/F_{p}$, where $F_{x}$,
$F_{y}$, $F_{z}$ and $F_{p}$ are defined in the following forms
(\citeauthor{mat95} \citeyear{mat95} and \citeauthor{moon02}
\citeyear{moon02}),
\begin{equation}
\label{fxyzp0} F_{x} = -\dfrac{1}{4\pi}\int B_{x}B_{z}dxdy,
\end{equation}
\begin{equation}
F_{y} = -\dfrac{1}{4\pi}\int B_{y}B_{z}dxdy,
\end{equation}
\begin{equation}
F_{z} = \dfrac{1}{8\pi}\int (B_{z}^{2}-B_{x}^{2}-B_{y}^{2})dxdy,
\end{equation}
\begin{equation}
\label{fxyzp1} F_{p} = \dfrac{1}{8\pi}\int
(B_{z}^{2}+B_{x}^{2}+B_{y}^{2})dxdy,
\end{equation}
are used to check the force-free condition of this quiet region. In
Equations (\ref{fxyzp0})-(\ref{fxyzp1}), $F_{x}$, $F_{y}$, $F_{z}$
are the components of the Lorentz force and $F_{p}$ is a
characteristic magnitude of the total Lorentz force. For this quiet
region (Figure 1B),
$F_{x}/F_{p}$, $F_{y}/F_{p}$ and $F_{z}/F_{p}$ are found to be 0.03,
-0.01 and -0.35, respectively. \citeauthor{mat95} (\citeyear{mat95})
argued that the magnetic fields with $\mid F_{z}/F_{p}\mid$ $\sim$
0.1 can be called force-free. \citeauthor{moon02}
(\citeyear{moon02}) studied 12 vector magnetograms and obtained $\mid F_{z}/F_{p}\mid$
ranging from 0.06 to 0.32 with a median value of 0.13. They
concluded that the photospheric magnetic fields are not so far from
force-free. Although $\mid F_{z}/F_{p}\mid$ of this quiet region is
a little larger than those of the active regions studied by
\citeauthor{moon02} (\citeyear{moon02}), we tentatively apply the
force-free extrapolation to reconstruct the magnetic fields above
the quiet region

Since the extrapolated fields are approximate solutions of the
force-free equations, their degree of force- and divergence-freeness
should be checked first. \citeauthor{whe00} (\citeyear{whe00})
introduced the criterion of force-freeness $\sigma_{J}$,
\begin{equation}
\label{sigma_J}
 \sigma_{J} = \dfrac{\sum_{i}J_{i}\sigma_{i}}{\sum_{i}J_{i}},
\end{equation}
where
\begin{equation}
\sigma_{i} = {\rm sin} \theta_{i} = \dfrac{\mid \textbf{J} \times
\textbf{B} \mid_{i} }{J_{i}B_{i}},
\end{equation}
and the criterion of divergence-freeness $f_{i}$,
\begin{equation}
\label{f_i}
 f_{i} = \dfrac{{\int_{\bigtriangleup S_{i}} \textbf{B}
 \cdot \textbf{dS}}}{{\int_{\bigtriangleup S_{i}} \mid\textbf{B}
 \mid\textbf{dS}}} \thickapprox \dfrac{(\nabla\cdot \textbf{B})_{i}
 \bigtriangleup\textbf{V}_{i}}{B_{i}A_{i}},
\end{equation}
to assess the quality of the extrapolations. $\sigma_{J}$ indicates
the weighted average of the sine of angle between the current
density and magnetic field. The average value of the magnitude of
$f_{i}$ is used to check if the system is close to divergence free,
where $A_{i}$ is the surface area of the small volume. The values of
$\sigma_{J}$ and $\langle|f_{i}|\rangle$ should be equal to zero, if
the force- and divergence-freeness of extrapolated field are fully
satisfied.

Table~\ref{T-bbb0} gives the values of $\sigma_{J}$ and $\langle
|f_{i}|\rangle$ in the four extrapolated fields. The maximum value
of Lorentz force ($F_{max}$ = max($\mid$ \textbf{J}$_{i}$ $\times$
\textbf{B}$_{i}$ $ \mid$), where \textbf{J}$_{i} = \nabla \times$
\textbf{B}$_{i}$) and the ratio of the total magnetic energy of the
NLFF extrapolated fields to that of the potential field
($\varepsilon$ = $E_{NLFF}/E_{pote}$) in the total volume are also
given in the Table~\ref{T-bbb0}. It can be found that these four
extrapolated fields meet the force- and divergence-freeness
basically and the orders of magnitude of these criteria are the same
for all the extrapolated fields. It is also found that the potential
field is computed with the highest degree of consistency, because
$\sigma_{J}$ and $\langle |f_{i}| \rangle$ of the potential field
are all smaller than those of three NLFF fields. For the NLFF
extrapolated fields, the divergence-freeness of the optimization
field is the best among these three NLFF fields, because $\langle
|f_{i}| \rangle$ in the optimization field is smaller than those in
the other two NLFF fields. The AVI field is reconstructed with the
best force-freeness, which can be seen from the valuse of $
\sigma_{J}$ and $F_{max}$ of these NLFF fields. $\varepsilon$ is
1.98, 2.10 and 2.81 for optimization, DBIE and AVI extrapolated
fields, respectively, which indicate that these NLFF fields are
reconstructed with a high degree of non-potentiality since the
values of $\varepsilon$ deviate strongly from unity.

Figure 2 shows the magnetic field lines of these four extrapolated
fields, where the red lines are the closed field lines and the blue
ones the open field lines (here the term of open is used to
characterize the field lines that leaves through the upper or
lateral boundary of the extrapolation box). It is found that the
topological structures of these extrapolated magnetic field lines
are similar on the whole, but the open magnetic field lines of the
NLFF fields are more than those of the potential field. To see the
distributions of the magnetic field lines clearly and to find their
differences, a green square region is cut near the disk center shown
in Figure 2, and the magnetic field lines in the region are shown in
Figure 3. It can be seen that the open magnetic field lines tend to
be located at the strong field region basically. A part of the
closed magnetic field lines of the NLFF fields can reach higher
altitudes than those of the potential field, and these closed lines
are more vertical than the potential one. Note that the extending
trends of the open magnetic field lines of the optimization field
are similar to those of the potential field. The reason may be that
the initial condition of the optimization field is a potential
field, but there are no such initial conditions for the AVI and DBIE
methods. In figure 3, it can also be seen that on the whole, the
distributions of the magnetic field lines of the AVI and DBIE fields
are similar, especially for closed magnetic field lines located at
the lower heights.

To compare these extrapolated fields, horizontal cuts of the four
extrapolated fields at three layers (z = 109, 545 and 2180 km) are
shown in Figure 4, where their similarities and differences are
displayed evidently. On the whole, the LOS components of these
extrapolated fields look similar and the expansion amplitudes of
these fields are almost the same. On the other hand, the transverse
magnetic fields are different, especially for their azimuths. For
the AVI and DBIE extrapolated fields, the arrows of the transverse
fields run across the strong LOS field regions as the height
increases, e.g. region A labeled in the third row. It also can be
found that though some fine features are different, the similarities
still exist between the transverse magnetic fields extrapolated with
the DBIE and AVI methods. For the potential and optimization fields,
the arrows of the transverse fields are inclined to converge in the
strong negative polarity region and diverge in the strong positive
polarity region.

In Figure 5, we plot the averages of the absolute strengthes of each
component of the extrapolated fields at different heights, where the
averages are calculated over each horizontal plane at the
corresponding height. It can be found that the strength of the
optimization field approaches  evidently that of the potential field as
the height increases. Note that as a whole, the profiles
of these extrapolated field strengthes are similar and the
extrapolated field strengthes are inclined to be the same values as
the height increases. It can be seen that $B_{x}s$ and $B_{y}s$ of
DBIE field are smaller than those of the potential field when the
height is low, but they exceed those of the potential field as the
height increases. For $Bzs$, the profiles of these field strengthes
are similar specially. It is noticed that the magnetic field
strength decreases quickly/slowly as the height is below/above
$\sim$ 1000 km. This trend is consistent for all the extrapolated
fields. To further compare the strength differences of the NLLF
fields ($B_{NLFF}$) and the potential field ($B_{P}$)
quantitatively, the averages of $\vert B_{NLFF}-B_{P}\vert$ for the
three NLFF extrapolated fields at the corresponding height are given
in Table~\ref{T-bbb}, where $ B_{NLFF}$ and $B_{P}$ are the total
field strength of a given position in different horizontal planes.
It can also be seen from Table~\ref{T-bbb}, the amplitudes of the
NLFF fields approach to that of the potential field as the height
increases. The differences of the magnetic field strength between
the NLFF fields and the potential field are of the order of 5 G on
the photosphere, and this order drops to 0.1 G when the height
reaches to $\sim$ 2000 km. The differences between the optimization
field and the potential field approach zero as the height is above
$\sim$ 4000 km.

In order to study the azimuth differences between the NLFF fields
and the potential field, the probability density functions (PDF) of
the shear angles at layers of z = 0, 545, 1090... 4905 km are
plotted. In this paper, the shear angles is defined as the average
over horizontal planes of the absolute value of the azimuth
difference between the NLFF and the potential fields at different
height. Figures 6, 7 and 8 show the PDFs of the shear angles of the
optimization, DBIE and AVI fields, respectively. In Figure 6, for
the optimization field, the shear angles decrease evidently with
height, and the mean value (the average of the absolute values over
each horizontal plane) changes from 58$^{\circ}$ at $z = 0$ km to
0$^{\circ}$ as $z >$ 4000 km. For the DBIE field in Figure 7, the
PDF profiles do not change evidently as the height increases. The
mean values of the shear angles increase with the height up to $z$
$\sim$ 2725 km, and decrease again above. In Figure 8, the mean
values of the shear angles of the AVI field increase slowly with
height, and the mean value of the shear angles reaches
72.2$^{\circ}$ at $z$ = 4905 km. While the PDF profiles do not
change significantly as $z >$ 2180 km.

In Figure 9, the distributions of the shear angles shown by the
contour lines are overlaid on the grey-scale map of the LOS magnetic
fields at three low layers of z = 0, 109, 545 and 1090 km. The red
contours are 60$^{\circ}$, 80$^{\circ}$ and 100$^{\circ}$, and the
blue ones -60$^{\circ}$, -80$^{\circ}$ and -100$^{\circ}$. Although
the mean values (the average of the absolute values) of the shear
angles are different for those extrapolated fields, the
distributions of the shear angles are very similar at the low
layers, especially for DBIE and AVI methods. It can be found that
the larger shear angles often appear near the edge of the strong
vertical magnetic field. It also can be found that the magnetic
field in the quiet Sun is inclined to be non-potential since there
are evident shear angles on the photosphere (z = 0 km).

\section{Discussions and Conclusions}
     \label{S-Discussion}

Although theories and models have been proposed to describe the
chromospheric and coronal magnetic field in the quiet Sun, it is
still an open problem to know the true chromospheric and coronal
magnetic fields. The magnetic field extrapolation is an alternative
method to study the chromospheric and coronal magnetic fields,
however there are limitations for applying the force-free
extrapolation to the quiet Sun.

The application of magnetic field extrapolation to measured
vector magnetograms of the active regions has been done with some
success recently, which are based on two considerations: First, the
reconstructed magnetic field can be assumed to be essentially
force-free; Second, sunspots with extremely high flux concentrations
are present in the vector magnetograms, so that the large-scale,
coronal magnetic structures linked to them might be possibly
determined by the dominant, largely force-free magnetic field only.
On the other hand, for the quiet Sun the usage of the magnetic field
extrapolation is less justified: the force-free assumption is not
satisfy completely (the plasma $\beta$ may be not low enough in the
quiet Sun) and the magnetic flux in the quiet regions is
concentrated on scales of the order of (or even smaller than) the
spatial spatial resolution in the normal magnetogram observations.

However, Kilogauss magnetic field has been found in the quiet Sun
recently ($e.g.,$ \citeauthor{sten73} \citeyear{sten73} and
\citeauthor{sten09} \citeyear{sten09}). Thus, the plasma may locally
satisfy the condition of $\beta$ $<1$ and the force-free assumption
is met approximatively, which, however, is a problem worthy to
investigate continually. Some authors have assumed the force-free
requirement is satisfied in the quiet Sun and used LFF field
extrapolation to study the properties of magnetic field in the quiet
Sun ($e.g.,$ \citeauthor{tu05}, \citeyear{tu05};
\citeauthor{regnier08} \citeyear{regnier08} and \citeauthor{zhao09},
\citeyear{zhao09}). In our work, we also assume that the force-free
requirement is satisfied in the quiet Sun and attempt to apply the
NLFF magnetic field extrapolation to the quiet Sun. A quiet region
with the high resolution observed by {\it Hinode} SOT/SP is chosen
as the boundary to extrapolate the magnetic field up to $\sim$ 5500
km above the photosphere. Four magnetic field extrapolation methods
are used to study the magnetic fields in the quiet Sun, and the
differences among these extrapolated fields are studied in detail.

In order to characterize the differences between the NLFF fields and
the potential field in the quiet Sun, we analyze the magnetic field
strengthes and the transverse azimuths of the NLFF fields and the
potential field. It is found that the field strength differences
between the NLFF fields and the potential field decrease evidently
as the height increases, although the amplitudes of these decrease are
different among these three NLFF fields. It is found that when the
height reaches to $\sim$ 2000 km, there are no evident
differences between the NLFF fields and the potential field, since
these differences are of the order of 0.1 G.

In the study of the transverse field azimuths, it is found that the
photospheric magnetic field is non-potential because there are
evident shear angles on the photosphere, the mean value of the shear
angles is about 58$^{\circ}$ on the photosphere. As the height
increases, the changes of shear angles are different among these
three NLFF fields. The shear angle tends to decrease rapidly as the
height increases for the optimization field, it increases gradually
as the height increases for the AVI field, and it stays practically
constant for the DBIE field. It is found that the larger shear
angles often appear near the edge of the strong vertical magnetic
fields, which are found to be very similar for three NLFF fields.

\begin{acks}
We thank the anonymous referee for helpful comments and suggestions.
{\it Hinode} is a Japanese mission developed and launched by
ISAS/JAXA, collaborating with NAOJ as a domestic partner, NASA and
STFC (UK) as international partners. Scientific operation of the
{\it Hinode} mission is conducted by the {\it Hinode} science team
organized at ISAS/JAXA. This team mainly consists of scientists from
institutes in the partner countries. Support for the post-launch
operation is provided by JAXA and NAOJ (Japan), STFC (U.K.), NASA,
ESA, and NSC (Norway). This work was partly supported by the
National Natural Science Foundation of China (Grant Nos.
10611120338, 10673016, 10733020, 10778723,
11003025 and 10878016), Important Directional Projects of Chinese
Academy of Sciences (Grant No. KLCX2-YW-T04) and National Basic
Research Program of China (Grant No. 2006CB806301).
\end{acks}

\begin{table}

\caption{The values of $\sigma_{J}$, $\langle |f_{i}| \rangle$,
$F_{max}$ and $\epsilon$ for the potential and three NLFF extrapolated fields.}
\label{T-bbb0}
\begin{tabular}{cccccccccccccccccccccc}
\hline                   
Method     & $\sigma_{J}$ & $\langle |f_{i}| \rangle$ & $F_{max}$ &$\varepsilon$\\
        &(Rad)    & 1$\times 10^{-4}$  &($G^{2}M^{-1}$)$\times$ $10^{-11}$    &  \\
\hline
Pote.          &0.50  &0.06   &0.55    &1.00   \\
Opti.           &0.97    &0.23   &1.21    &1.98    \\
DBIE           &0.92    &1.01   &0.92    &2.01     \\
AVI            &0.89  &2.93   &0.45    &2.81      \\
\hline
\end{tabular}
\end{table}

\begin{table}
\caption{The average of $\vert B_{NLFF}-B_{P}\vert$ for three NLFF
extrapolated fields at different heights. } \label{T-bbb}
\begin{tabular}{cccccccccccccccccccccc}
\hline   
Method     &\multicolumn{5}{c}{$\vert B_{NLFF}-B_{P}\vert$         (G)} \\
\hline
                            & & &Height (km)     \\

                &z = 0     &z = 1090   &z = 2180  &z = 3270 &z = 4360 \\

\hline
Opti.           &4.6713   &0.0591      &0.0091    &0.0007  &0.0001    \\
DBIE           &4.6713    &0.5490      &0.0748    &0.1544  &0.1628    \\
AVI            &4.6713    &0.5226      &0.1068    &0.0895  &0.1162     \\
\hline
\end{tabular}
\end{table}

 \begin{figure}
    \centerline{\includegraphics[width=1.\textwidth,clip=]{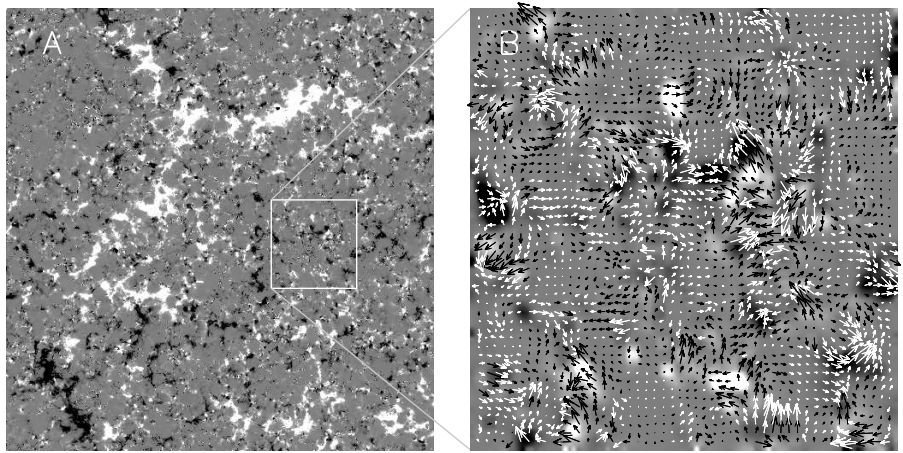}}
    \caption{Left (A): line-of-sight (LOS) magnetic field observed from 00:23 UT to 01:48 UT on
 April 16, 2007. Right (B): the vector magnetogram of the region marked in the left panel with
 the white rectangle. The background image is the LOS magnetic field and the arrows stand for the
 transverse magnetic field.}
   \end{figure}
\begin{figure}
   \centerline{\includegraphics[width=1.\textwidth,clip=]{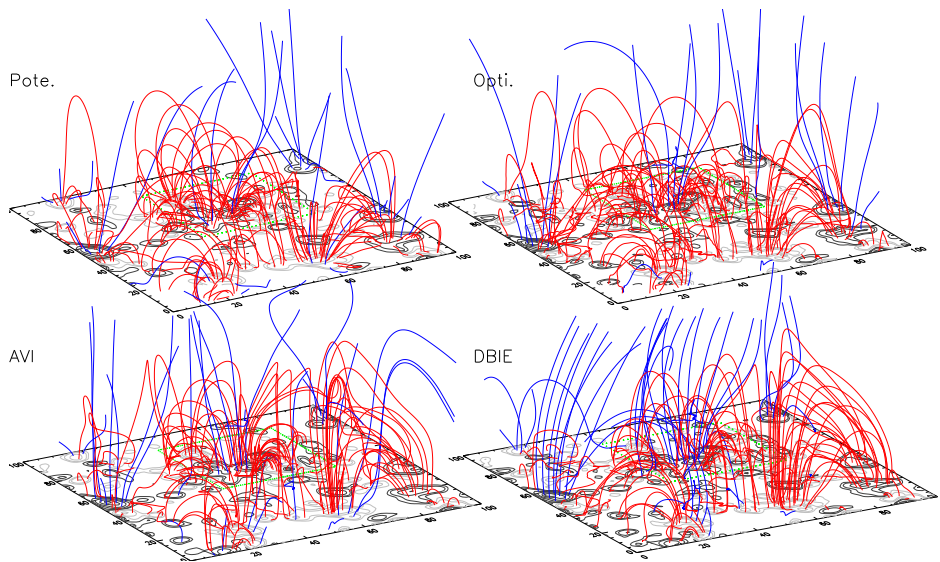}}
   \caption{The magnetic field lines for three NLFF fields
   (Opti, AVI and DBIE) and a potential field.}
\end{figure}

\begin{figure}
   \centerline{\includegraphics[width=1.2\textwidth,clip=]{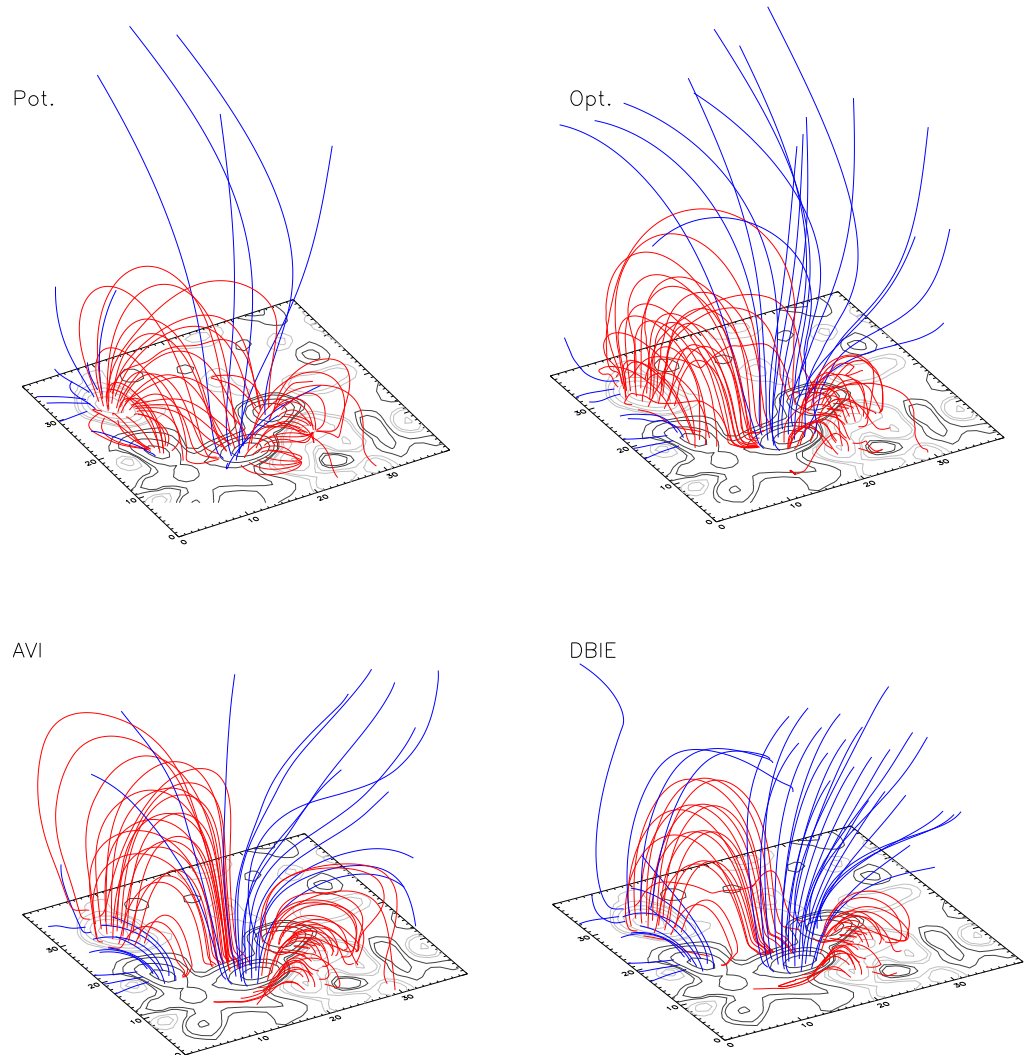}}
   \caption{The magnetic field lines for three NLFF fields
   (Opti, AVI and DBIE) and a potential field, strarting within
   the sub-region labeled by a green square in Fig 2.}
\end{figure}

\begin{figure}
\label{mag3}
   \centerline{\includegraphics[width=1.0\textwidth,clip=]{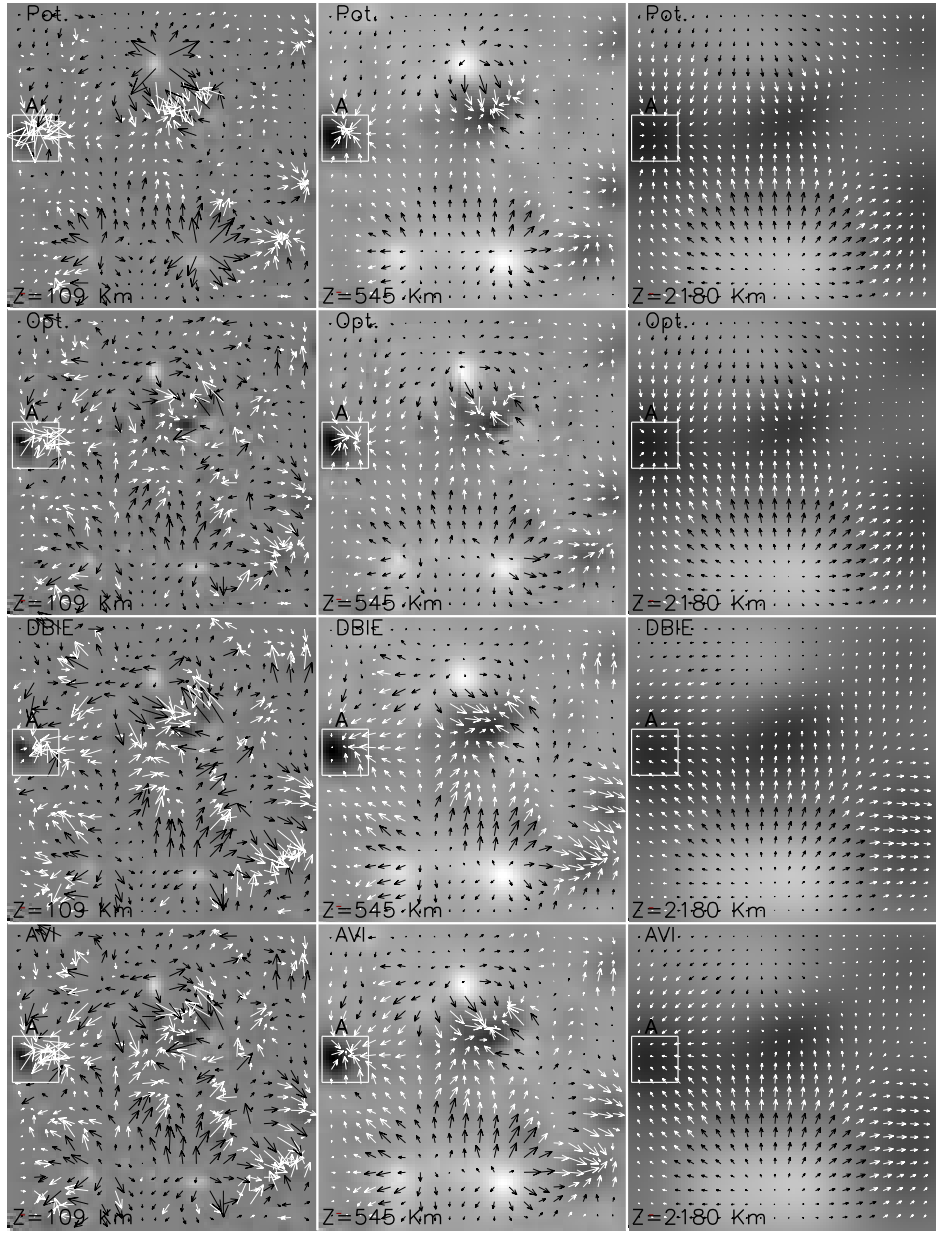}}
   \caption{Vector magnetograms of extrapolated field at the heights of
   z = 109, 545 and 2180 km. The background image is the LOS magnetic field and the
   black and white arrows stand for transverse magnetic field. The rows
   1, 2, 3 and 4 correspond to the potential field, three NLFF fields
   extrapolated with optimization, DBIE and AVI method, respectively.
   The columns 1, 2 and 3 show different heights.}
\end{figure}

\begin{figure}
   \centerline{\includegraphics[width=1.2\textwidth,clip=]{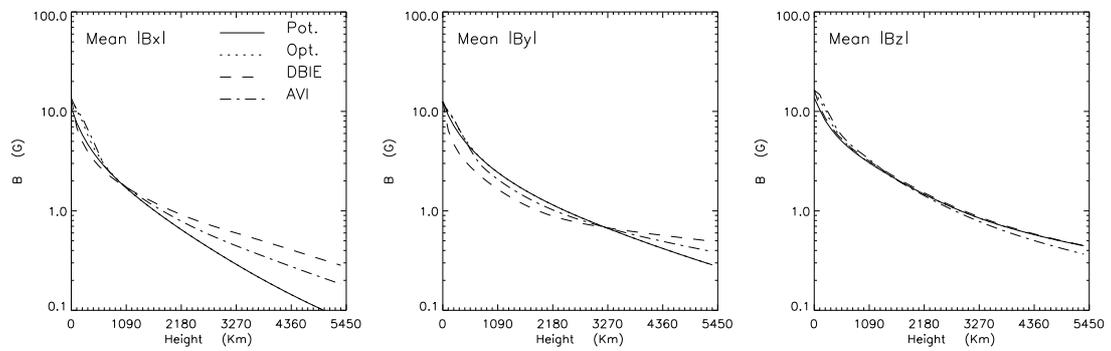}}
   \caption{The average of absolute strengths of $B_{x}$, $B_{y}$
   and $B_{z}$ of the extrapolated magnetic field versus the height,
   the different style lines indicate the extrapolation methods used.}
\end{figure}

\begin{figure}
\label{s_pdf}
   \centerline{\includegraphics[width=1.\textwidth,clip=]{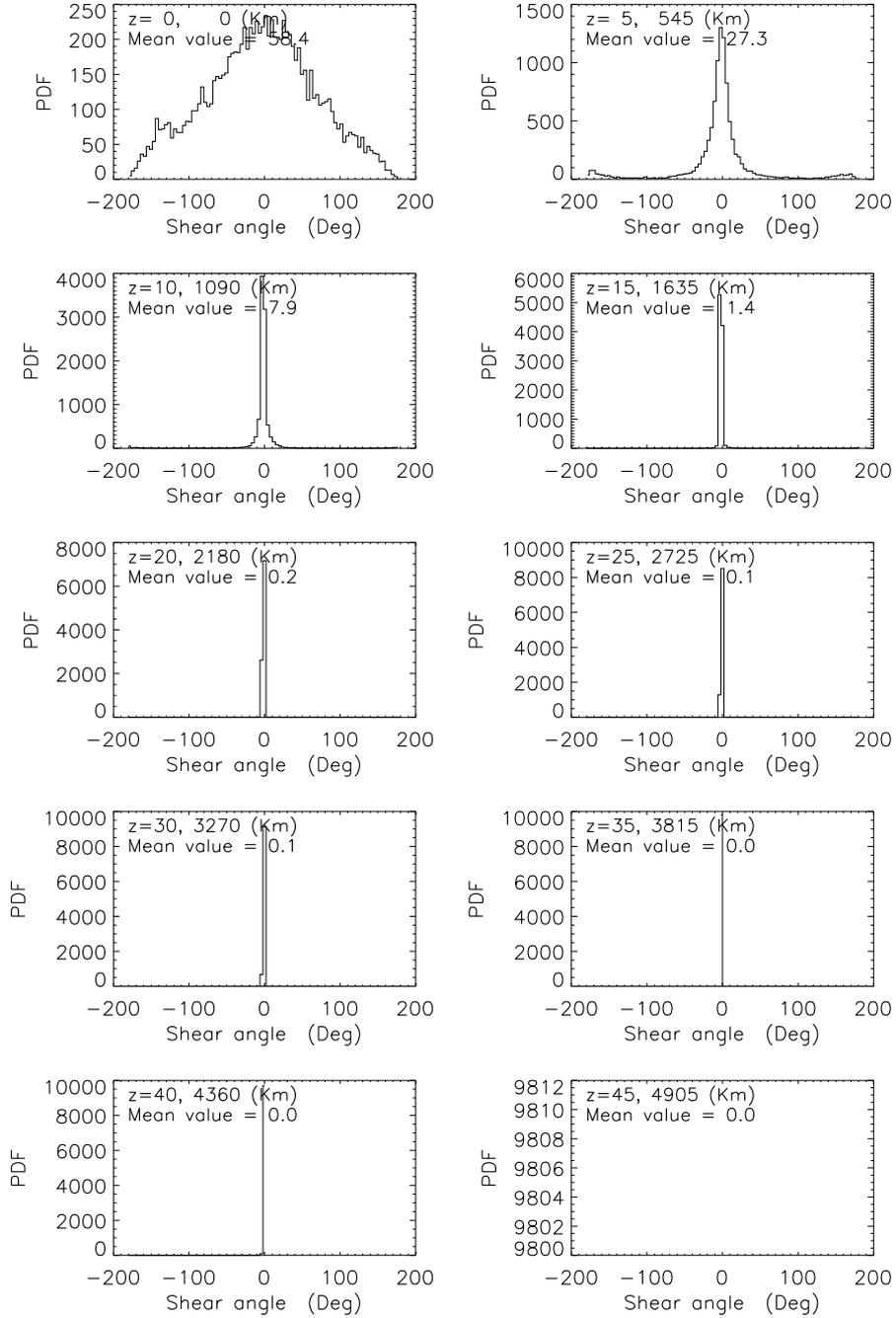}}
   \caption{The PDF of the shear angles at different layer (z = 0, 545, 1090...
   4905 km) for optimization method. Mean value plotted in each panel is the
   average of the absolute value of the shear angle.}
\end{figure}

\begin{figure}
   \centerline{\includegraphics[width=1.\textwidth,clip=]{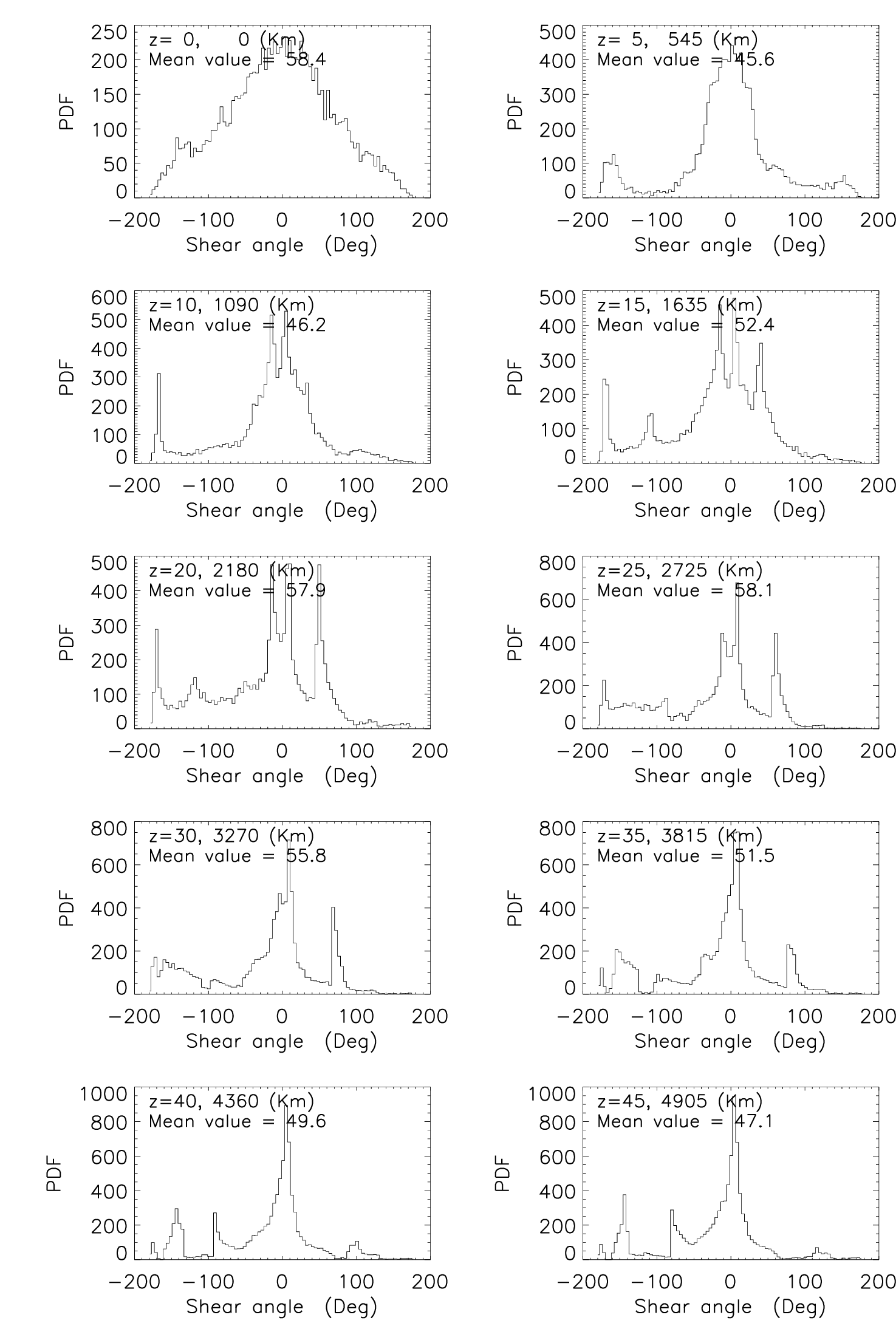}}
   \caption{Same as Figure 6 but for DBIE method.}
  \end{figure}

\begin{figure}
   \centerline{\includegraphics[width=1.\textwidth,clip=]{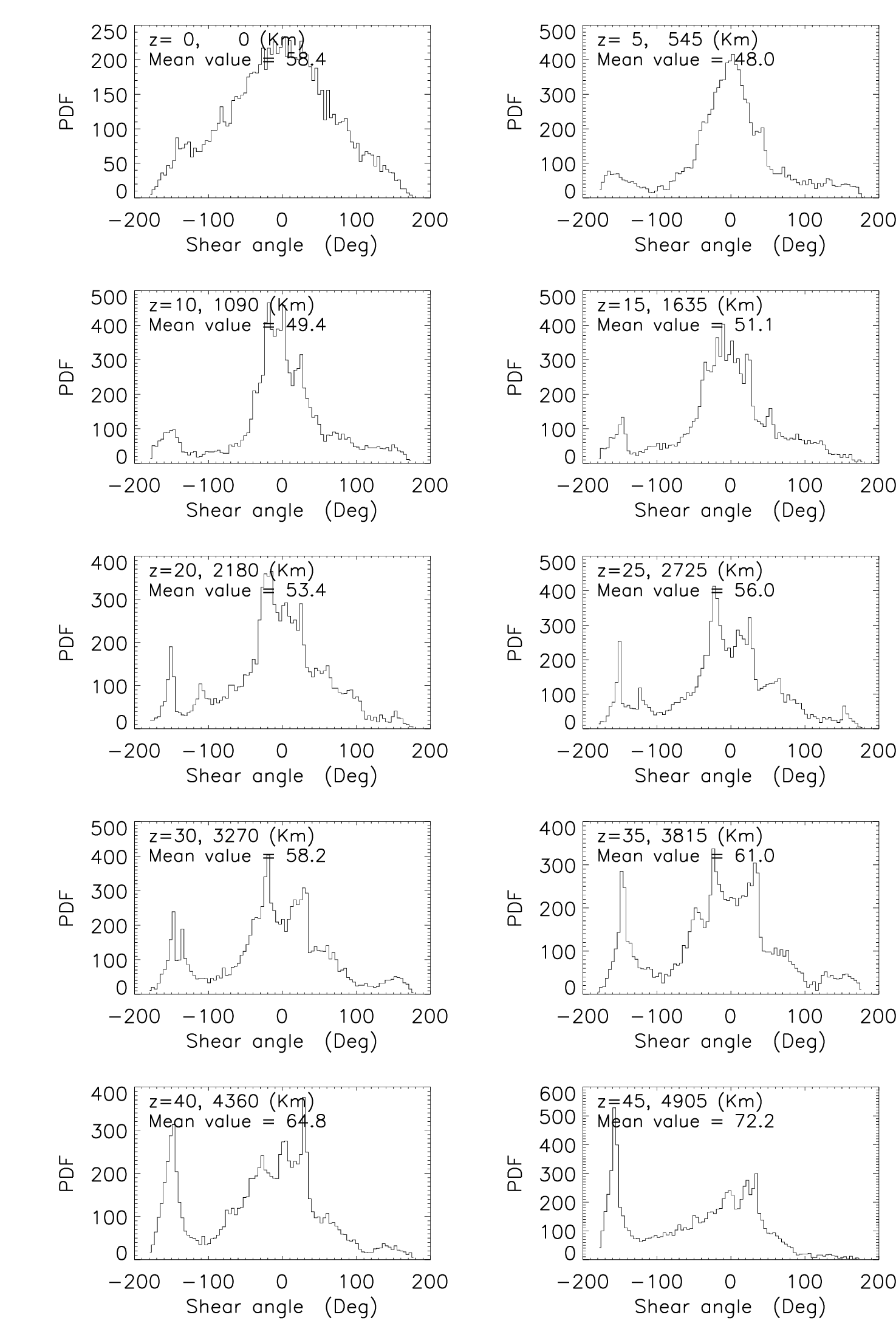}}
   \caption{Same as Figure 6 but for AVI method.}
  \end{figure}

\begin{figure}
   \centerline{\includegraphics[width=1.\textwidth,clip=]{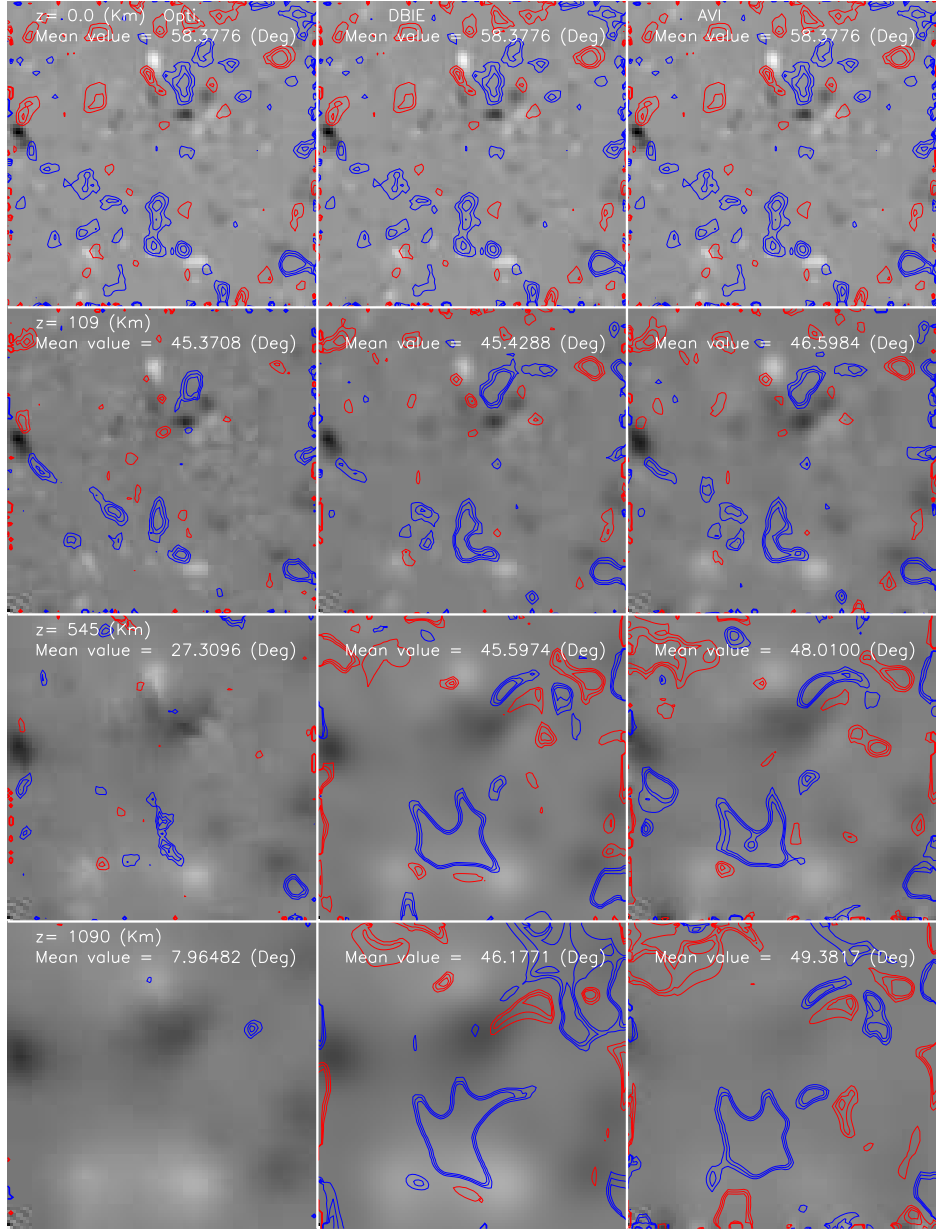}}
    \caption{The distributions of the shear angles at different layer
     (z = 0, 109, 545 and 1090 km) for three NLFF extrapolated fields.
     The contours are $\pm$60$^{\circ}$, 80$^{\circ}$, 100$^{\circ}$
     and red/blue contours represent positive/negative values.
     The columns 1, 2 and 3 correspond to the optimization,DBIE and AVI
     method, and the rows 1, 2 3,and 4 are for z = 0, 109, 218 and 327 km,
     respectively. The mean value reported in each panel is the average of the absolute value of the shear angle.}
  \end{figure}
\end{article}
\end{document}